\documentclass[seceq]{ptptex}

\usepackage[dvips]{graphicx}
\usepackage{times,txfonts,bm}

\notypesetlogo                       

\title{Similarity of the gravitational contributions\\ to anomalous magnetic moment
}

\author{Konosuke Sawa}

\inst{Department of Physics, Tokyo University of Science, Tokyo
162-0861, Japan
}

\abst{
A precise measurement of a muon $(g-2)$ experiment recently reported a deviation from the Standard Model. This
implies the need to an extension of the model and suggests underlying mechanisms. We
hypothesize that the influence of higher dimensional gravity causes this
$(g-2)$ deviation. This effect would appear as Kaluza-Klein modes and
Nambu-Goldstone modes on a $3$-brane. These classical effects produce
new contributions to the magnetic moment. In this paper, we perform a
consistency check by comparing the results derived from the Born and
Schr{\"o}dinger approximations. This comparison reveals a similarity
between KK-dilaton modes and the brane stretching effect.  
}

\begin{document}

\maketitle

\section{Introduction}
Recently, the BNL E821 group reported a  precise measurement of 
the muon anomalous magnetic moment. Based on their result, we
obtained the new average value $a^{(\textrm{exp})}_{\mu}=11659208(6)\times 10^{-10}
(\pm 0.7 \textrm{ppm})$ \cite{Brown:2000sj}. This is inconsistent with the theoretical
value, $a^{(\textrm{SM})}_{\mu}=11659182(6)\times 10^{-10} (\pm0.7 \textrm{ppm})$,
taken from H{\"o}cker et al. \cite{Hocker:2004xc}. The existence of this
discrepancy, $\Delta a_{\mu}\equiv a^{(exp)}_{\mu}-a^{(SM)}_{\mu}=(26\pm
9.4)\times10^{-10}$, reveals that the SM is not realized strictly within
the low-energy regime when the difference exceeds the calculation uncertainties for
the Hadronic process and experimental error. Although we can consider
various reasons for this deviation, we focus on the influences of extra
dimensions (see also \citen{Casadio:2000pj}). In this paper, we consider
a braneworld scenario
\cite{Arkani-Hamed:1998rs,Antoniadis:1998ig,Arkani-Hamed:1998nn,Sundrum:1998sj,Sundrum:1998ns}
in which the SM particles are confined to the $3$-brane world volume $M_4$
and only gravity can propagate freely in the bulk
space-time $M_4\times B$ with $(4+n)$-dimensions, where $M_4$ is the
$4$-dimensional Minkowski space-time and $B$ is
a given compact manifold. The bulk space-time $M_4\times B$ produces KK
modes and brane fluctuations. These fields always couple with all the
matter fields similarly to gravity. The implication of this interaction is that the
stress-energy tensor of the matter fields is the source of the KK modes
and brane fluctuations. This suggests the KK modes and brane fluctuations
are closely related to the energy scale of the physical process. From this
point of view, we study the deviation of the anomalous magnetic moment due
to classical effects. 

This paper is organized as follows. In \S \ref{Kaluza-Klein approach},
we treat the contribution of bulk gravity on the basis of the Kaluza-Klein
theory. In \S \ref{Higher dimensional relativistic approach}, we investigate
the effect of the brane fluctuations, and in \S \ref{Conclusions} we
summarize the paper. The notation we use for the indices of the coordinates is
summarized in Table \ref{Summary of our notation}. 

\begin{table}[t]
   \begin{center}
   \caption{Summary of notation}
         \begin{tabular}{lccc}\hline\hline
       &$0,1,\cdots,n+3$    &\quad$0,1,\cdots,3$&\quad $4,5,\cdots,n+3$    \\ \hline
Curved     &$M,N,\cdots$  &$\mu,\nu,\cdots$ &$m,n,\cdots$   \\ \hline
Local Lorentz      &$A,B,\cdots$   &$\alpha,\beta\cdots$&  \\ \hline
       \end{tabular}
       
\label{Summary of our notation}
              \end{center}
\end{table}

\section{Kaluza-Klein approach}\label{Kaluza-Klein approach}
\begin{figure}[t]
\centerline{
\includegraphics[height=3cm,width=3cm]{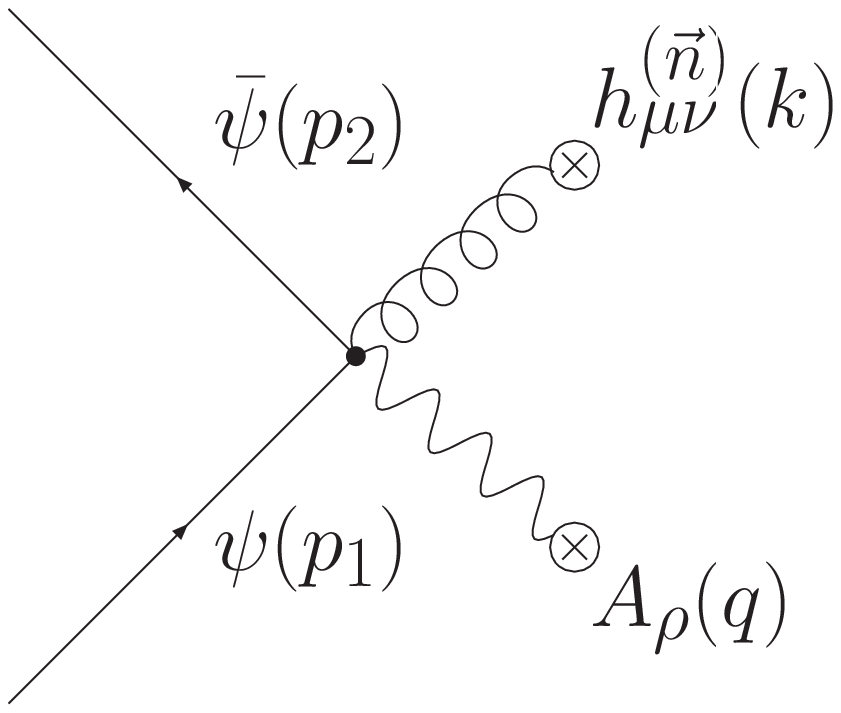}
\hspace{1mm}
\raisebox{13.5mm}{\includegraphics[height=0.3cm,width=0.3cm]{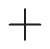}}
\hspace{1mm}
\includegraphics[height=3cm,width=3cm]{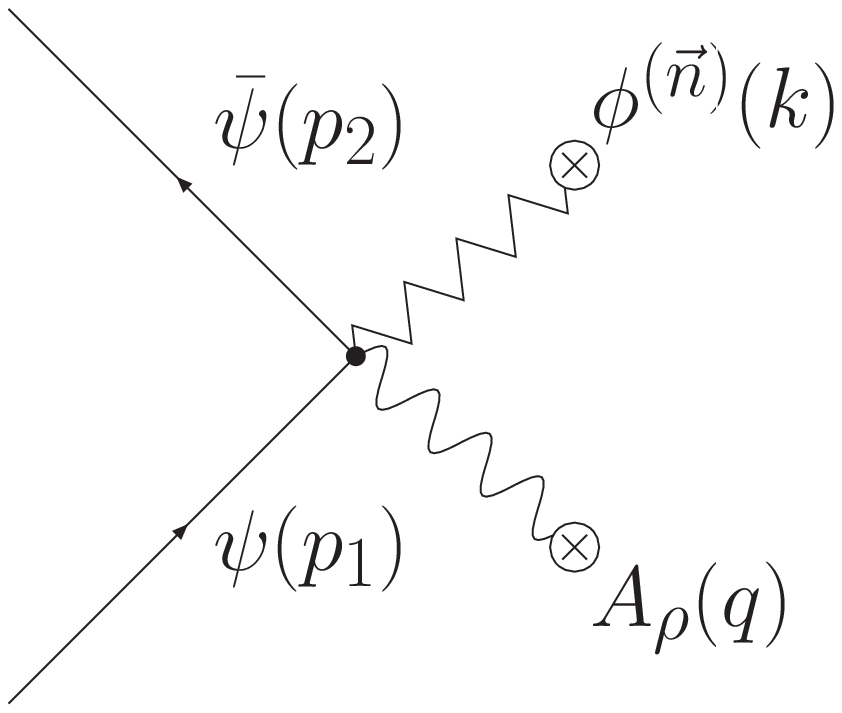}
}
\caption{Feynman diagrams for an anomalous magnetic moment in bulk gravity.}\label{figure}
\end{figure}

We investigate the anomalous magnetic moment originating from the
KK modes. In the following, we assume that the configuration
of the extra dimensions possesses on $n$-torus with a common radius, which determines
the masses of the KK modes. Although this assumption is unrealistic, it
simplifies the treatment, both conceptually and computationally. 
Further, when we carry out the KK-reduction procedure, the
$(4+n)$-dimensional Fierz-Pauli action decomposes into a KK-tower of
gravity, $(n-1)$ gauge fields, and $n(n-1)/2$ scalar components
\cite{Han:1998sg}. 
Using the prescription \cite{Graesser:1999yg}
\begin{align}
\kappa h_{\mu\nu}\Rightarrow {\kappa} \sum_{\vec{n}}(h^{(\vec{n})}_{\mu\nu}(x)-\omega\eta_{\mu\nu}\phi^{(\vec{n})}(x))\label{graesser prescription},
\end{align} 
we can easily determine the new effects of bulk gravity, as in the case of $4$D
gravity, where $\omega=\sqrt{(n-1)/3(n+2)}$, ${\kappa}=\sqrt{16\pi G_{N}}$.
$G_{N}$ is Newton's constant, which is related to Plank mass $M_{Pl}$ as $M_{Pl}^{2}=(8\pi G_{N})^{-1}$.
The field $h^{(\vec{n})}_{\mu\nu}$ and
$\phi^{(\vec{n})}=\phi^{(\vec{n})}_{mm}$ represent the KK-gravity and 
the KK-dilaton modes, respectively. The coupling of the KK modes and matter is given by
\begin{align}
{\cal L}_{\textrm{int}}=-\frac{{\kappa}}{2}\sum_{\vec{n}}\left\{ h^{\mu\nu(\vec{n})}T_{\mu\nu}+\omega\phi^{(\vec{n})}T^{\mu}_{\mu}\right\}\label{kk dilaton and kk gravity coupling},
\end{align}
where $T_{\mu\nu}$ is the conserved symmetric stress-energy tensor of
the matter fields. For simplicity, we regard the matter fields as
consisting of a $U(1)$ gauge field and a fermion. In this interaction,
it is noteworthy that the KK-dilaton modes couple with matter only
through the trace part, $T^{\mu}_{\mu}$. This indicates that the
behavior of the KK-dilaton modes differs from $4$D gravity and
KK-gravity. References \citen{Han:1998sg} and \citen{Giudice:1998ck} provide the Feynman rules for the couplings of the KK modes with matter.
Using those rules and the assumption that $h^{(\vec{n})}_{\mu\nu}$ and $\phi^{(\vec{n})}$ are static in time: 
\begin{align}
h^{(\vec{n})}_{\mu\nu}=\left(
  \begin{array}{cc}
    0   &  0  \\
    0   &  h_{ij}^{(\vec{n})}({\bm x}) \\
  \end{array}
\right), \quad
\phi^{(\vec{n})}=\phi^{(\vec{n})}({\bm x}),
\end{align}
where ${\bm x}$ represents the spatial coordinates $x^{i}$ $(i=1,2,3)$,
we can calculate a classical effect of the anomalous magnetic moment using the tree-level
KK modes. Additional diagrams that contribute to the magnetic moment are
shown in Fig. \ref{figure}. Thus, the amplitude can be written
\begin{eqnarray}
i{\cal M}&=&\bar{u}(p_{2})\sum_{\vec{n}}\Bigl\{ie\frac{{\kappa}}{4}(\eta_{\mu\rho}\eta_{\nu\sigma}+\eta_{\mu\sigma}\eta_{\nu\rho}-2\eta_{\mu\nu}\eta_{\rho\sigma})\gamma^{\sigma}A^{\rho}(q)h^{\mu\nu(\vec{n})}(k)\nonumber\\&&+ie\frac{3{\kappa}\omega}{2}\gamma^{\rho}A_{\rho}(q)\phi^{(\vec{n})}(k)\Bigr\}u(p).
\end{eqnarray}
When we take the non-relativistic limit for the fermions, and retain 
terms up to first order in the momentum, we can interpret ${\cal M}$ as a
potential well by using the Born approximation. Because the potential for a magnetic moment is   
\begin{align}
V({\bm x})=-g\frac{e}{2m}{\bm S}\cdot{\bm B}({\bm x}),
\end{align}
where ${\bm S}$ is the particle spin, $g$ is the Land{\' e}
$g$-factor, and ${\bm B}$ is a magnetic field, we obtain the anomalous
magnetic moment as
\begin{align}
\left(\frac{g-2}{2}\right)_{\textrm{B}}=\frac{{\kappa}}{2}\sum_{\vec{n}}\left\{h^{(\vec{n})}_{ii}({\bm
 x})+3\omega{\phi}^{(\vec{n})}({\bm x})\right\}\label{KK amplitude g-2}.
\end{align}
This shows that the KK-gravity and KK-dilaton modes give new contributions to the anomalous magnetic moment. 

Next, we examine a method for calculating the quantity $(g-2)$ using the Schr{\"o}dinger approximation.
The fermion action in conventional $4$D gravity is given by
\begin{align}
S=\int d^{4}x \sqrt{-g}\Bigl\{i{\bar \psi}e^{\mu}_{\alpha}\gamma^{\alpha}\left(\partial_{\mu}-ieA_{\mu}-\frac{1}{2}\omega^{\beta\gamma}_{\mu}\sigma_{(\beta\gamma)}\right)\psi-m{\bar \psi}\psi\Bigr\},
\end{align}
where
\begin{eqnarray}
\omega^{\alpha \beta}_{\mu}&=&\frac{1}{2}e^{\alpha \nu}(\partial_{\mu}e^{\beta}_{\nu}-\partial_{\nu}e^{\beta}_{\mu})+\frac{1}{4}e^{\alpha \nu}e^{\beta \sigma}(\partial_{\sigma}e^{\gamma}_{\nu}-\partial_{\nu}e^{\gamma}_{\sigma})e_{\gamma \mu}-(\alpha \leftrightarrow \beta).
\end{eqnarray}
The spin connection $\omega^{\alpha \beta}_{\mu}$ is essential to maintain the Lorentz and gauge symmetries of $\psi$ in gravity. We assume that the gravitational metric is static in time:
\begin{align}
g_{\mu\nu}=\left(
  \begin{array}{cc}
    1   &  0  \\
    0   &  \eta_{ij}+\kappa h_{ij} \\
  \end{array}
\right).
\end{align} 
By employing the Sch{\" o}dinger approximation using the metric (refer
to Appendix of \citen{Sawa:2005py}, which demonstrates this
approximation), we obtain the potential term for the magnetic moment:
\begin{align}
\frac{-ie}{2m}\sigma^{ij}{e}^{k}_{i}{e}^{l}_{j}F_{kl}=\frac{-e}{2m}\Bigl\{(2+{{\kappa}}h_{ii})\frac{\sigma^{i}}{2}H^{i}-\frac{\kappa}{2}\sigma^{j}{h}^{j}_{i}H_{i}\Bigr\}.\label{megnetic moment due to 4d gravity}
\end{align}
In general, the term proportional to $\sigma^{i}H^{i}/2$ represents the
magnetic moment. Using the prescription (\ref{graesser prescription}),
we can estimate the effect of bulk gravity based on the KK-theory to be
given by
\begin{align}
\frac{ie}{2m}\sigma^{ij}{e}^{k}_{i}{e}^{l}_{j}F_{kl}\Rightarrow& \frac{-e}{2m}\biggr[\Bigl\{2+{\kappa}\sum_{\vec{n}}(h^{(\vec{n})}_{ii}+3\omega\phi^{(\vec{n})})\Bigr\}\frac{\sigma^{i}}{2}H^{i}\nonumber\\&-\sum_{\vec{n}}\Bigl\{\frac{{\kappa}}{2}\sigma^{i}h^{i(\vec{n})}_{j}H_{j}-{\kappa}\omega\phi^{(\vec{n})}\frac{\sigma^{i}}{2}H^{i}\Bigr\}\biggr]\\
=&\frac{-e}{2m}\biggr[2+{\kappa}\sum_{\vec{n}}(h^{(\vec{n})}_{ii}+2\omega\phi^{(\vec{n})})\frac{\sigma^{i}}{2}H^{i}-\sum_{\vec{n}}\frac{{\kappa}}{2}\sigma^{i}h^{i(\vec{n})}_{j}H_{j}\biggr].\label{twisting effect in KK}
\end{align}
From this equation, we obtain the muon anomalous magnetic moment,
\begin{align}
\left(\frac{g-2}{2}\right)_{\textrm{S}}=\frac{1}{2}{\kappa}\sum_{\vec{n}}(h^{(\vec{n})}_{ii}+2\omega\phi^{(\vec{n})}).
\end{align}
This result shows that the trace part $h^{(\vec{n})}_{ii}$ is consistent
with Eq. (\ref{KK amplitude g-2}), while the KK-dilaton part
$\phi^{(\vec{n})}$ is not. It seems that the classical effect of the
KK-dilaton modes cannot be evaluated accurately by calculating the
scattering amplitude. The off-diagonal components of $4$D gravity,
$h_{ij}$, result in the last term of (\ref{megnetic moment due to 4d
gravity}), which does not affect $(g-2)$. However, using the
prescription (\ref{graesser prescription}), the KK-dilaton modes
introduce additional contribution to $(g-2)$ from the last term. 

\section{Brane fluctuations}\label{setup}
\label{Higher dimensional relativistic approach}
\begin{figure}[t]
\hspace{1mm}
\centerline{
\includegraphics[height=3.3cm,width=3.3cm]{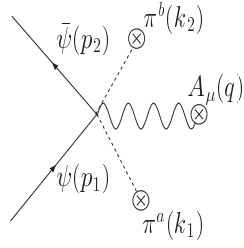}
}
\caption{Tree-level diagram that contributes to the anomalous magnetic moment due to the NG-mode.}\label{brane diagram}
\end{figure}

We now calculate the magnetic moment by use of brane fluctuations. The
existence of a brane causes a symmetry breaking of the bulk isometry,
and it introduces brane fluctuations regarded as Nambu-Goldstone (NG)
modes
\cite{Akama:1982jy,Sundrum:1998sj,Dobado:2000gr,Alcaraz:2002iu,Bando:1999di,Kugo:1999mf,Dvali:2000bz,Murayama:2001av,Hisano:1999bn}.
In accordance with the action principle, a brane is created with the
shape of $Y^{m}(x)$, whose possible solution $Y^{m}(x)=Y_{0}^m$ occupies
a certain point in B. We consider this to be the ground state of brane
fluctuations. We assume that our brane tension is negative, which
ensures the stability of NG-modes. When the bulk space-time includes a brane, the bulk metric possesses the isometric group
\begin{align}
G(M_{D})=G(M_{4})\times G(B).
\end{align}       
For simplicity, we assume a factorizable metric:
\begin{align}
G_{MN}=\left(
  \begin{array}{cc}
    {\eta}_{\mu\nu}(x)   & 0   \\
     0  &  -{\cal G}_{mn}(y)  \label{Gold stone metric ansatz}\\
  \end{array}
\right).
\end{align}
The induced metric \cite{Sundrum:1998sj,Sundrum:1998ns,Akama:1982jy} is described as
\begin{align}
g_{\mu\nu}(x)=G_{MN}\left(Y(x)\right)\partial_{\mu}Y^{M}\partial_{\nu}Y^{N}.
\end{align}
When the brane ground state $Y_{0}^m$ is created on $B$, zero-mode
excitations are produced. Therefore, the brane excitation along a broken
generator produces the zero-mode that corresponds to the NG-mode. The NG-mode is parametrized as \cite{Alcaraz:2002iu}
\begin{align}
Y^{m}(x)&=Y^{m}(Y_{0},\pi^{a}(x))\nonumber\\
&=Y^{m}_{0}+\frac{1}{f^2}\xi^{m}_{a}(Y_{0})\pi^{a}(x)+O(\pi^2) \label{GB parametrize},
\end{align}
where the killing vector $\xi^{m}_{a}(Y_{0})$ is normalized by the condition
\begin{align}
\xi^{m}_{a}(Y_{0})\xi^{n}_{b}(Y_{0}){\cal G}_{mn}=\delta_{ab}.\label{nomalization}
\end{align}
Thus, we can estimate the influence on $4$D physics by calculating the coupling of the SM particle with the NG-mode $\pi^{a}(x)$.
Using the gauge fix condition $Y^{\mu}(x)=x^{\mu}$ and the Eqs. (\ref{GB
parametrize}) and (\ref{nomalization}), the induced metric to be
\begin{align}
g_{\mu\nu}&=\eta_{\mu\nu}-\frac{1}{f^4}\partial_{\mu}\pi^{a}\partial_{\nu}\pi^{a}. \label{linearized induced metric}
\end{align}
Substituting the above expression into the minimal brane action with a
negative tension,
\begin{align}
 S_{\textrm{brane}}=\int d^4 x \sqrt{-g}\biggl\{-f^4+{\cal L}(g_{\mu\nu})\biggr\}\label{eq:1},
\end{align}
we obtain the effective brane action,
\begin{align}
S_{\textrm{brane}}=\int d^4 x \biggl\{-f^4+{\cal L}(\eta_{\mu\nu})+\frac{(\partial_{\mu}\pi^{a})(\partial_{\nu}\pi^{a})}{2}\left(\eta^{\mu\nu}+\frac{T^{\mu\nu}}{f^4}\right)\biggr\}.\label{interaction1}
\end{align}
Here, $f^4$ represents the brane tension and ${\cal L}$ is the matter
Lagrangian. The interaction terms for the fermion become  
\begin{align}
S_{\textrm{int}}=&\int dx^4 \frac{1}{2f^4}\partial^{\mu}\pi^{a}\partial^{\nu}\pi^{a}\Bigl\{-\eta^{\mu\nu}\left({\bar \psi}i\gamma^{\rho}D_{\rho}\psi-m{\bar \psi}\psi\right)\nonumber\\
&+\frac{1}{2}{\bar \psi}i\gamma^{\mu}D^{\nu}\psi+\frac{1}{2}{\bar \psi}i\gamma^{\nu}D^{\mu}\psi+\frac{\eta^{\mu\nu}}{2}\partial^{\rho}({\bar \psi}i\gamma_{\rho}\psi)\nonumber\\
&-\frac{1}{4}\partial^{\mu}({\bar \psi}i\gamma^{\nu}\psi)-\frac{1}{4}\partial^{\nu}({\bar \psi}i\gamma^{\mu}\psi)\Bigr\}.
\end{align}
The Feynman rules for the NG-mode interaction is given in Ref. \citen{Alcaraz:2002iu}. Using those rules, as in the previous section, we can calculate the contribution to the anomalous magnetic moment using the Born approximation. With the assumption that the NG-mode is static in time,
\begin{align}
\pi^{a}= \pi^{a}({\bm x}),
\end{align}
we obtain the diagram contributing to ($g-2$), which is shown in Fig. \ref{brane diagram}.
The additional amplitude is written
\begin{align}
i{\cal M}=&\frac{1}{2}\bar{u}(p_{2})\Biggl[\frac{-ie\delta^{ab}}{4f^4}\Bigl\{2\gamma^{\nu}k_{2\nu}k_{1\mu}+2\gamma^{\nu}k_{1\nu}k_{2\mu}-4\gamma_{\mu}(k_{1}\cdot k_{2})\Bigr\}\Biggr]\nonumber\\
&\times u(p_{1})A^{\mu}(q)\pi^{a}(k_{1})\pi^{b}(k_{2}).\label{eq:2}
\end{align}
From the third term, we obtain the following additional term for the
anomalous magnetic moment:
\begin{align}
\left(\frac{g-2}{2}\right)_{\textrm{B}}=\frac{1}{2f^4}\partial^{i}\pi^{a}\partial^{i}\pi^{a}.
\end{align}
From this point, we calculate the anomalous magnetic moment using the Schr{\"o}dinger approximation. Assuming that $\pi^{a}$ is static in time, we substitute the induced metric
\begin{align}
g_{\mu \nu}=\left(
  \begin{array}{cc}
    1   & 0   \\
    0   & \eta_{ij}-\frac{1}{f^4}\partial_{i}\pi^{a}\partial_{j}\pi^{a}   \\
  \end{array}
\right).
\end{align}
Therefore, the induced vierbein and its inverse are given by
\begin{align}
e^{i}_{k}&=\delta^{i}_{k}-\frac{1}{2f^4}\partial^{i}\pi^{a}\partial_{k}\pi^{a}\\
e^{k}_{i}&=\delta^{k}_{i}+\frac{1}{2f^4}\partial^{k}\pi^{a}\partial_{i}\pi^{a},
\end{align}
respectively. Then, using the Schr{\"o}dinger approximation, we obtain the potential term
\begin{align}
\frac{ie}{2m}\sigma^{ij}e^{k}_{i}e^{l}_{j}F_{kl}=\frac{-e}{2m}\biggl[\biggl\{2+\frac{\partial^{i}\pi^{a}\partial^{i} \pi^{a}}{f^4}\biggr\}\frac{{\sigma}^{j}}{2}{H^{j}}-\frac{\sigma^{i}}{2}\frac{\partial^{i}\pi^{a}\partial^{j} \pi^{a}}{f^4}{H^{j}}\biggr].\label{energy shift NG-mode}
\end{align}
Because the coefficients of the part proportional to
$\frac{\sigma^{i}}{2}{H^{i}}$ represent the magnetic moment, we obtain
the following anomalous magnetic moment:
\begin{align}
\left(\frac{g-2}{2}\right)_{\textrm{S}}=\frac{1}{2f^4}\partial^{i}\pi^{a}\partial^{i}\pi^{a}.
\end{align}
The NG-mode approach is consistent with both the Born and
Schr{\"o}dinger approximations. However, it is possible to regard the
classical field $\pi^{a}$ as the brane stretching effect at low energy
\cite{Sawa:2005py}. The existence of a brane separates bulk space-time
into a $4$-dimensional Minkowski space-time and an $n$-dimensional extra
space. Static brane fluctuations allow extra dimensional coordinate
$Y^{m}$ to acquire $x^{i} (i=1,2,3)$ dependence through the
parametrization (\ref{GB parametrize}). Because the NG mode originates
from bulk gravity, based on dimensional analysis, the physics are
conjectured to be characterized by the order of the fundamental Plank mass,
$M_f$. On the other hand, the interaction term of (\ref{interaction1})
indicates that the stress-energy tensor of matter fields is a source of
the NG-modes. This implies that the NG-mode should be closely related to the energy
scale $E$ of the physical process on brane. Thus, if we introduce the
dimensionless coordinate $Ex^i$, the NG mode would be parametrized as 
\begin{align}
\pi^{a}=M_{f}Ex^{i}{\tilde e}^{a}_{i},\label{brane stretching effect}
\end{align} 
where the basis vectors
\begin{align}
\frac{\partial Y^{m}}{\partial x^{i}}=\frac{EM_{f}}{f^2}\xi^{m}_{a}{\tilde e}^{a}_{i}
\end{align}
satisfy the completeness relation
\begin{align}
{\cal G}_{mn}\frac{\partial Y^{m}}{\partial x^i}\frac{\partial
 Y^{n}}{\partial x^j}=\frac{E^{2}M_{f}^{2}}{f^4}\delta_{ij},
\end{align}
i.e.,
\begin{align}
\delta_{ab}{\tilde e}^{a}_{i}{\tilde e}^{b}_{j}=\delta_{ij}\label{completeness}
.
\end{align}
where $\tilde{e}^a_i$ are the basis vectors for the spatial part of
Minkowski space-time. We can easily verify that 
Eq. (\ref{brane stretching effect}) satisfies the equation motion $\delta
S/\delta \pi^{a}=0$ 
derived from the action (\ref{interaction1}). This
solution is valid for $E\ll M_f$, because we are considering the low
energy expansion of action (\ref{eq:1}). Using Eqs. (\ref{brane
stretching effect})-(\ref{completeness}), all the terms in
Eq. (\ref{eq:2}) and the second and third terms in (\ref{energy shift
NG-mode}) result in additional contributions to the
$g$-factor. Collecting all of these, we obtain the results:
\begin{align}
\left(\frac{g-2}{2}\right)_{\textrm{B}}=\frac{E^2 M_{f}^2}{2f^4},\quad
\left(\frac{g-2}{2}\right)_{\textrm{S}}=\frac{E^2 M_{f}^2}{f^4}.
\end{align}
This discrepancy indicates that the influence of the brane stretching
 effect is similar to that of the KK-dilaton modes.
Also, if we consider the case $M_{f}\sim f$ with a TeV scale, we would
 obtain a contribution of the appropriate order for the muon $(g-2)$ for the
 muon scale $E \approx 106$ $[\textrm{MeV}]$. 

\section{Summary}\label{Conclusions}
We have studied the classical effects of the KK modes and brane
fluctuations on the magnetic moment. Employing KK-theory, we observe
that KK-gravity and KK-dilaton modes contribute positive quantities to
the Land{\'e} $g$-factor, which would be a desirable contribution to the
muon ($g-2$) value. Then, we found that the brane fluctuations also
contribute positive quantities. In both studies, we used two different
approximations -- the Born and Schr{\"o}dinger approximations. Although
we know that the two approximations yield the same results for SM particles,
we find that the Born approximation cannot accurately evaluate the
KK-dilaton modes and brane-stretching effect. 
This indicates that the classical KK-dilaton modes and brane-stretching
effect produce novel behavior that differs from that of the conventional $4$D fields.




\section*{Acknowledgements}
This study was partly supported by Iwanami F{\= u}jyukai.

\bibliography{apssamp}

\end{document}